\documentclass[twocolumn,showpacs,amsmath,amssymb]{revtex4}
\usepackage{graphicx}
\usepackage{color}

\begin{document}
%\title{Among the adiabatic and the sudden limit: The coupling between the quantum breathing and the quantum friction}
%\title{Analogy of RKKY oscillation in cold atom: the energy inserting in non-adiabatic process}
\title{Analogy of RKKY oscillations to the heat exchange in cold atoms}
\author{Ching-Hao Chang and Tzay-Ming Hong}
\affiliation{Department of Physics, National Tsing Hua University, Hsinchu, Taiwan 300, Republic of China}
\date{\today}

\begin{abstract}
An oscillatory term is found in both the energy expectation and dynamics of a wave-packet in a time-varying harmonic trap and infinite potential well. 
They are proved to oscillate in coherence with the time lapse within each period depending on both  the cutoff in transition energies and the specific route via which the potential is being varied.
This oscillatory term is general to arbitrary potential forms since it derives from the interference between crossed transition trajectories.
Close analogy is made to the Ruderman-Kittel-Kasuya-Yosida interaction for giant-magnetoresistance trilayers, where
many-body quantum interference among scattering states renders the oscillation as a function of spacer width.
This connection reveals the generality of quantum friction due to parasitic oscillations. 
\end{abstract}

\maketitle

\section{Introduction}

The adiabatic and sudden processes are two opposite
limits in the time-dependent quantum mechanics. 
%When the walls move adiabatically slow, the phase space is conserved
%and the instantaneous state population remains the same. In contrast, the shape of the wave-packet in the sudden limit is conserved, but the
%population is changed, except in the pathological case of a contracting infinite well. 
The adiabatic tuning is involved in experiments, such as the advanced
cooling technique\cite{cooling-tech}, quantum
computing\cite{computing1,computing2}, and the state broadening in
optical lattices\cite{optical-lattice}. The combination of
adiabatic and sudden processes is also used to minimize the transition time between different thermal equilibrium states\cite{bang-bang}. Although analytic forms conveniently exist for these two limits,
cares need to be taken when identifying them to real systems\cite{adi-the}. 
How to describe the physics for a time-varying potential when neither limit is applicable remains a challenging task. 
Interesting phenomena have been proposed to exist in this regime, 	for instance, 
the parasitic oscillations that give rise to the quantum friction when the Hamiltonians at different time do not commute, 
as is the case when the potential profile changes at finite rates\cite{q-f1,q-f2}.

\begin{figure}
\includegraphics[width=0.5\textwidth]{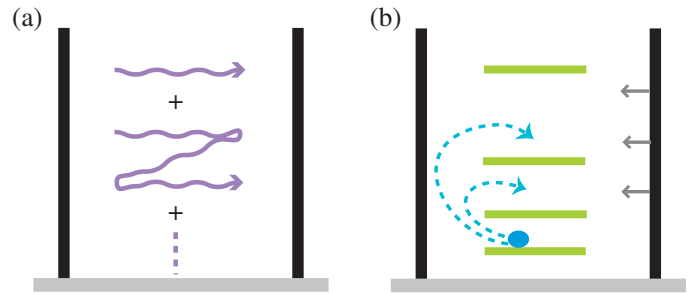}
\caption{(color online) Interference diagrams for  (a) the RKKY oscillation and (b) oscillatory term in the energy expectation in a contracting potential well. The curve lines denote scattering paths with different rounds
between the interfaces. The solid lines represent energy levels and the dotted lines are transition paths.} 
\label{fig-rkky}
\end{figure}

%Unlike the friction that lead to dissipations, the concept of quantum friction is used to describe the oscillatory term in the energy expectation or, more specifically, why this term can be negative while the potential trap is contracting. 
Unlike the friction that lead to dissipations, the quantum frictional process is reversible and energies can be stored in or taken out of the oscillations\cite{bang-bang}. 
Non-adiabatic processes allow the transition between different instantaneous states, which leads to the parasitic oscillations. Ostensibly it seems difficult to deduce any general property in these systems without the knowledge of the specific paths of transition.
However, since the parasitic oscillations are a collective behavior that results from the contribution of all trajectories, this task has been shown to be possible. 
One of the most famous examples in many-body systems is the carrier-mediated interaction between magnetic impurities or thin films, namely, the Ruderman-Kittel-Kasuya-Yosida (RKKY) coupling\cite{rkky}.
This coupling is dominated by the quantum interference between scattering paths\cite{md} of the mediating electrons and the resulting strength oscillates as a function of separating distance
with a period of half Fermi wavelength.
Similar to the RKKY coupling, \emph{the dynamic parasitic oscillations mediate the coupling} between boundaries. While the energy input from the moving boundaries plays the role of coupling strength, the oscillation is now in time instead of space. If we take as an example the giant magnetoresistance (GMR)\cite{gmr} system in which the RKKY mechanism has successfully applied, the close resemblance of interference diagrams for RKKY and the time-varying potential trap is exemplified in Fig.\ref{fig-rkky}.

In Section II we first examine the harmonic trap because the time-dependence of its basis lies solely in the frequency $\omega(t)$. 
The expectation values of energy and position-momentum correlation that represent the parasitic oscillations are calculated analytically for an arbitrary wave-packet. 
We extend our study in Section III  to an infinite well to demonstrate that these oscillations and the time lapse within each period they share is calculable are general.
The interesting analogy between RKKY oscillation and the energy input from the time-varying trap is arranged in the final Section IV, together with some discussions and conclusions. 
By use of formulae in Section II, we calculate the energy expectation of harmonic oscillators varying with a constant adiabatic parameter in Appendix A, and reproduced routes that have been reported\cite{mu} to be frictionless. 
%In Appendix A,  formulae in Section II is used to examine the frictionless routes in the oscillator with constant adiabatic parameter. This is to connect with the technique\cite{mu}.
Appendix B provides calculations to explain why the oscillation ceases to exist in the adiabatic and sudden limits.

\section{Harmonic trap}
\subsection{Energy and Position-Momentum Correlation}
We consider an one-dimensional time-dependent harmonic trap,
$\hat{H}(t)=\hat{p}^{2}/2m+m\omega^{2}(t)\hat{x}^{2}/2$, with an
initial angular frequency $\omega(0)=\omega_{0}$. An orthonormal basis can be constructed out of the time-dependent
eigenfunctions\cite{lohe}
\begin{align}
\notag
\Psi_{n}(t,x)&=\Big(\frac{m\omega_{0}}{\pi\hbar}\Big)^{1/4}~\frac{e^{-i(n+1/2)\int_{0}^{t}dt'(\omega_{0}/b^{2})}}{(2^{n}b\cdot n!)^{1/2}}\\
&\times
e^{i(m/2\hbar)(\dot{b}/b+i\omega_{0}/b^{2})x^{2}}H_{n}\Big[\Big(\frac{m\omega_{0}}{\hbar}\Big)^{1/2}\frac{x}{b}\Big],
\label{eq:wavefunc}
\end{align} 
where
$b(2\hbar/m\omega_{0})^{1/2}=bL_{0}$ is the time-dependent
length scale with the scaling function, $b(t)$, satisfying
\begin{align}
\ddot{b}+\omega^{2}(t)b=\omega^{2}_{0}/b^{3},
\label{eq:b-omega}
\end{align}
and the initial conditions $b(0)=1$.

To calculate the energy expectation, we first use  $\hat{H}(t)$ to operate on the basis function
\begin{align}
\notag &\hat{H}(t)|\Psi_{n}(t)\rangle
=e^{-i(n+1/2)\int_{0}^{t}dt'(\omega_{0}/b^{2})}e^{i(m/2\hbar)\dot{b}\hat{x}^{2}/b}\\
&\times\Big[(n+\frac{1}{2})\frac{\hbar\omega_{0}}{b^{2}}+\frac{m}{2}(\dot{b}^{2}-b\ddot{b})(\frac{\hat{x}}{b})^{2}
+\frac{\dot{b}}{2b}\{\hat{x}, \hat{p}\}\Big]|\phi_{n}(b)\rangle,
\label{eq:h-operate}
\end{align}
where $|\phi_{n}(b)\rangle$ is the normalized instantaneous
eigenfunction for
$(\hat{p}^{2}/2m+m\omega_{0}^{2}\hat{x}^{2}/2b^{4})$ with eigenvalue $(n+1/2)(\hbar
\omega_{0}/b^{2})$.

Given any wave-packet $\psi(0,x)=\psi_{0}(x)$, its evolution follows 
$\psi(t,x)=\sum_{n}c_{n}\Psi_{n}(t,x)$ with the
projection
\begin{align}
c_{n}=\langle\Psi_{n}(t=0)|\psi_{0}\rangle=\langle
\phi_{n}(b=1)|e^{-i(m/2\hbar)\dot{b}(0)x^{2}}\psi_{0}\rangle.
\end{align}
By use of Eq.(\ref{eq:h-operate}), the energy expectation can be obtained as
\begin{align}
\notag
\langle\hat{H}(t)\rangle
%=&\sum_{nl}c_{n}c^{*}_{l}\langle\Psi_{l}(t)|\hat{H}(t)|\Psi_{n}(t)\rangle\\
=&\sum_{n,l}e^{-i (n-l)\int_{0}^{t}dt'(\omega_{0}/b^{2})}\\
\notag
&\langle e^{-im\dot{b}(0)\hat{x}^{2}/2\hbar}\psi_{0}|\phi_{l}(1)\rangle\langle\phi_{l}(b)|\\
\notag&\Big[(n+\frac{1}{2})\frac{\hbar\omega_{0}}{b^{2}}+\frac{m}{2}(\dot{b}^{2}-b\ddot{b})(\frac{\hat{x}}{b})^{2}
+\frac{\dot{b}}{2b}\{\hat{x}, \hat{p}\}\Big]\\
&|\phi_{n}(b)\rangle
\langle\phi_{n}(1)|e^{-im\dot{b}(0)\hat{x}^{2}/2\hbar}\psi_{0}\rangle.
\label{eq:packet-energy0}
\end{align}
Since our choice of time-dependent basis is not the eigenfunctions of energy operator, the energy expectation in Eq.(\ref{eq:packet-energy0}) contains both diagonal and  non-diagonal elements.
The non-diagonal elements are non-zero only when $n$ and $l$ share the same parity; namely, $n-l$ is an even number.

The two summations in Eq.(\ref{eq:packet-energy0}) can be removed by appealing to the identity operator after we change the length scale in the second inner product
from $b$ to $1$ and reexpress the $(n-l)$ in the exponent by the Hamiltonian operator. Equation (\ref{eq:packet-energy0}) now becomes
\begin{align}
\notag
\langle \hat{H}(t) \rangle=&\langle\psi_{0}|e^{im\dot{b}(0)\hat{x}^{2}/2\hbar}e^{i\frac{\hat{H}_{0}}{\hbar}\int_{0}^{t}dt'/b^{2}}\\
\notag&\Big[\Big(\frac{\hat{H}_{0}}{b^{2}}-\frac{m}{2}b\ddot{b}\hat{x}^{2}\Big)+\frac{m}{2}\dot{b}^{2}\hat{x}^{2}+\frac{\dot{b}}{2b}\{\hat{x},\hat{p}\}\Big]\\
&e^{-i\frac{\hat{H}_{0}}{\hbar}\int_{0}^{t}dt'/b^{2}}e^{-im\dot{b}(0)\hat{x}^{2}/2\hbar}|\psi_{0}\rangle,
\label{eq:packet-energy1}
\end{align}
with $\hat{H}_{0}=\hat{p}^{2}/2m+m\omega^{2}_{0}\hat{x}^{2}/2$.
Being independent of $\dot{b}$, the first term in the square bracket can be identified as the energy in the adiabatic limit:
\begin{align}
\frac{\hat{H}_{0}}{b^{2}}-\frac{m}{2}b\ddot{b}\hat{x}^{2}=\frac{-1}{2m}\Big(\frac{\partial}{b\partial x}\Big)^{2}+\frac{m}{2}\omega^{2}(t)b^{2}\hat{x}^{2},
\end{align}
where Eq.(\ref{eq:b-omega}) has been used.
%which represents the average energy for the rescale wave-packet $\psi_{0}(b(t)x)$ in the trap $\omega(t)$; namely, the potential energy. 
In the mean time, the ${m}\dot{b}^{2}\hat{x}^{2}/2$ term describes the extra kinetic energy due to the pull or push by the moving potential.
The third term can be easily shown to be
%\textquotedblleft$\frac{\dot{b}}{2b}\{\hat{x},\hat{p}\}$" which
proportional to $[\hat{H}(0), \hat{H}(t)]$, which gives rise to the \emph{quantum friction}
\cite{q-f1,q-f2} when the Hamiltonians at different times do not commute. The reason is that the positive/negative sign of  $\langle\{\hat{x}, \hat{p}\}\rangle$ reflects the expansion/contraction motion of the
wave-packet, while $\dot{b}/b$ keeps track of similar motions of the trap.
If the wave-packet moves along with the trap, it receives a positive work of ${\dot{b}}\langle\{\hat{x},\hat{p}\}\rangle/{2b}$. On the other hand, against intuitions, the packet can sometimes resist and behave oppositely, which causes the work to turn negative and diminishes its energy. If the initial wavefunction consists of only one $\psi_0$ Eq.(\ref{eq:wavefunc}) as, Eq.(\ref{eq:packet-energy1}) can be solved straightforwardly as
\begin{align}
\langle\hat{H}(t)\rangle_{n}=\frac{\big(2n+1\big)\hbar}{4\omega_{0}}\Big(\frac{\omega_{0}^{2}}{b^{2}}+\omega^{2}b^{2}+\dot{b}^{2}\Big).
\label{eq:eigenE}
\end{align}
The friction term expectedly becomes zero because the eigenfunction always moves coherently with the trap.

For a general $\psi_0$, Eq.(\ref{eq:packet-energy1}) can be simplified by using 
(1) ladder operators in $\hat{H}_{0}$, $a_{0}$ and $a_{0}^{+}$, and 
(2) the commutation relations
\begin{align}
\notag
a_{0} e^{\gamma a_{0}^{+}a_{0}}&=e^{\gamma}e^{\gamma a_{0}^{+}a_{0}}a_{0},\\
\notag
a_{0} e^{\eta (a_{0}^{+}+a_{0})^{2}}&=e^{\eta
(a_{0}^{+}+a_{0})^{2}}\big[a_{0}+2\eta(a_{0}^{+}+a_{0})\big)].
\end{align}
By use of both relations,
Eq.(\ref{eq:packet-energy1}) can be simplified to
\begin{align}
\notag\langle \hat{H}(t) \rangle
=&\hbar\Big(\frac{\omega_{0}}{b^{2}}+\frac{1}{2\omega_{0}}\big(\dot{b}^{2}-b\ddot{b}\big)\Big)\Big[\frac{1}{2}+\langle\psi_{0}|\Lambda^{+}\Lambda|\psi_{0}\rangle\Big]\\
\notag
&+\hbar\mathrm{Re}\Big[\Big(\frac{1}{2\omega_{0}}\big(\dot{b}^{2}-b\ddot{b}\big)-i\frac{\dot{b}}{b}\Big)\\
&\times
e^{-2i\int_{0}^{t}dt'\omega_{0}/b^{2}}\langle\psi_{0}|\Lambda\Lambda|\psi_{0}\rangle\Big]
\label{eq:oscillationE}
\end{align}
where the new ladder operators, $\Lambda$ and $\Lambda^{+}$, are obtained from the old set via the Bogoliubov transformation,
\begin{align}
\Lambda\equiv a_{0}-i\frac{\dot{b}(0)}{2\omega_{0}}(a_{0}^{+}+a_{0}).
\label{eq:Lambda}\end{align} 
Another interesting property pertinent to the dynamics is the expectation value for the position-momentum
correlation operator $\hat{C}=\{\hat{x},\hat{p}\}$,
%and Lagrangian $\hat{\mathcal{L}}(t)=\hat{p}^{2}/2m-m\omega^{2}(t)\hat{x}^{2}/2$:
\begin{align}
\notag\langle\hat{C}\rangle=&\hbar\frac{b\dot{b}}{\omega_{0}}\Big[\frac{1}{2}+\langle\psi_{0}|\Lambda^{+}\Lambda|\psi_{0}\rangle\Big]\\
&+\hbar\mathrm{Re}\Big[\big(\frac{b\dot{b}}{\omega_{0}}-i\big)e^{-2i\int_{0}^{t}dt'\omega_{0}/b^{2}}\langle\psi_{0}|\Lambda\Lambda|\psi_{0}\rangle\Big].
\label{eq:oscillationC}
\end{align}
The derivations so far depend on our selection of a time-varying trap that avoids generating a complex phase factor, $\omega_{0}=\sqrt{\omega^{2}(0)+\ddot{b}(0)}$. 
One example is the trap with a constant adiabatic parameter $\mu\equiv \dot{\omega}/\omega^{2}$.
The energy expectation was found to switch from being sinusoidal to hyperbolic as $|\mu|\ge 2$ by Rezek {\it et al.}\cite{mu}. 
Furthermore, they discovered the oscillation to be intimately linked to the existence of frictionless routes when $|\mu|< 2$.
Detailed discussion are arranged in Appendix A where identical results are rederived by our approach, i.e., Eq.(\ref{eq:oscillationE}).
From now on, we shall confine ourselves to initial conditions $\ddot{b}(0)=0$ that guarantees $\omega_{0}=\omega(0)$ to be real.
 
It may seem surprising for a system as simple as the harmonic potential within the single-particle picture that the energy expectation in Eq.(\ref{eq:oscillationE}) 
should contain an oscillatory term which, within one period, requires
\begin{align}
2\int_{0}^{T}dt'\omega_{0}/b^{2}(t)=2\pi. \label{eq:period}
\end{align}
where $T$ is in general time-dependent and thus is not a proper period.
To understand its origin, let us go back and examine the mathematic structure in Eq.(\ref{eq:packet-energy0}). We have already shown  that only the even-$(n-l)$ terms need to be considered. Therefore, the $T$ shared by Eq.(\ref{eq:oscillationE}) and Eq.(\ref{eq:oscillationC}) is the least common multiple of these different phases as defined by Eq.(\ref{eq:period}).
%As was stated in Eq., the different phases of odd terms cancel of survive to contribute to the oscillations
% with different phases, $(-(n-l)\int dt'(\omega_{0}/b^{2}))$, tend to cancel out except when $n-l$ equals even integers.
%Note that the wave-packet evolves in the time-dependent basis $\Psi_n (t,x)$ in Eq.(\ref{eq:wavefunc}), while the energy expectation is determined by the instantaneous eigenstates $\phi_n (t,x)$.
%These two states are only equivalent in the adiabatic limit. The dynamical phase value appears and evolves with time in the expectation value because of the time-dependent level crossing in the instantaneous eigenstates. The phase term that respected to lowest-energy transition is most important term to decide the final frequency in the energy oscillation.

%Before extending the calculation to the infinite potential well, we firstly discuss the interplay between parasitic oscillations and energy in non-adiabatic process.

%The fluctuation of energy arises from the quantum interference between transition trajectories.
%Physically, we believe this arises from the coupling between the quantum breathing and
%the quantum friction, which shall be elaborated in more details in the next section. 

\subsection{Coupling between quantum friction and quantum breathing}
The  parasitic oscillations give rise to the breathing effect\cite{breathing} which term was originally proposed to describe the large-amplitude fluctuations of wave-packet caused by the excess energy\cite{optimal-exp} when a harmonic trap was expanded suddenly. In this case, the energy expectation of the final state is a constant. Only the position-momentum correlation $\langle C\rangle$ contains the oscillatory term with frequency, $1/T$, determined solely by the frequency $\omega$ characterizing the final trap. In the following, we shall allow the trap to contact at a general rate. For a medium  rate $(b\dot{b}/\omega_{0})\sim 1$, the quantum friction will enter and render the quantum breathing with the oscillation period sensitively affected by the specific route via which the final stage is reached.

\begin{figure}
\includegraphics[width=0.4\textwidth]{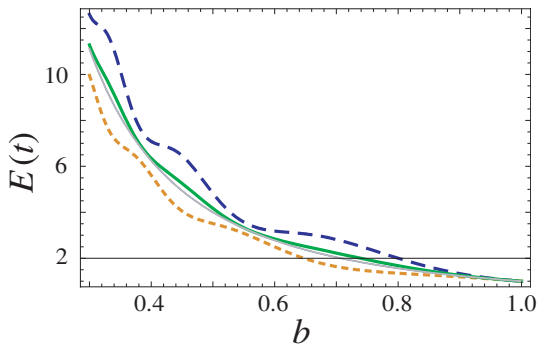}
\caption{(color online) The energy expectation in Eq.(\ref{eq:oscillationE2}) is plotted as a function of $b$ for a linear contraction with
$\dot{b}=-\omega_{0}/4$ and $\ddot{b}=0$ (thick solid line), and the adiabatic
process with $E(b)=E_{0}/b^{2}$(thin solid line). In comparison, the coherent state with $\{x_{0},p_{0}\}$=$\{\sqrt{\hbar/m\omega_{0}},-\sqrt{\hbar m\omega_{0}}\}$ that moves away from the contracting boundary is 
plotted in dashed line and the state with the opposite motion for $\{x_{0},p_{0}\}$=$\{\sqrt{\hbar/m\omega_{0}},\sqrt{\hbar m\omega_{0}}\}$ is plotted in dotted line. From now on, the energies in all figures are expressed in unit of the initial energy.
%The energy scale $E_{0}$ is defined as $\hbar \omega_{0}/2$, which is the energy expectation before modulating the potential.
} 
\label{fig-energy}
\end{figure}

We start with a linearly contracting harmonic trap, where $\dot{b}$ is constant, to pave ways for our later extension to other potential profiles in Section III. 
Set the initial wavefunction to lie in the ground state $|\phi_0(b=1)\rangle$
and insert it to Eq.(\ref{eq:oscillationE}) give the
energy  expectation at any later time
\begin{align}
\notag E(t)=&\hbar\Big(\frac{\omega_{0}}{b^{2}}+\frac{\dot{b}^{2}}{2\omega_{0}}\Big)\Big[\frac{1}{2}+\big(\frac{\dot{b}}{2\omega_{0}}\big)^{2}\Big]\\
\notag
&-\frac{\hbar\dot{b}^{2}}{2\omega_{0}}\Big[\big(\frac{\dot{b}}{2\omega_{0}}\big)^{2}+\frac{1}{b}\Big]\cos\big(2\frac{\omega_{0}}{b}t\big)\\
&+\frac{\hbar\dot{b}^{2}}{2\omega_{0}}\frac{\dot{b}}{2\omega_{0}}\Big(\frac{1}{b}-1\Big)\sin\big(2\frac{\omega_{0}}{b}t\big).
\label{eq:oscillationE2}
\end{align}
For a mild contraction, $(b\dot{b}/\omega_{0})\sim 1$, the oscillation terms
are as important as the first term in Eq.(\ref{eq:oscillationE2}). Besides demonstrating the oscillatory feature,  Fig.\ref{fig-energy} also shows a higher energy expectation than the adiabatic result which is an artifact of our choice to start from the ground state. Had we chosen a mixed state such as the coherent state exp($i p_{0} \hat{x}/\hbar$) exp($-ix_{0}\hat{p}/\hbar$) $|\phi_0(1)\rangle$,
where $x_{0}$/$p_{0}$ denotes the initial position/momentum of the state, their relative size could be reversed.
For the coherent state that moves toward the contracting boundary, the energy modulation is smaller than the adiabatic result because of the reduction from the quantum friction. 

\begin{figure}
\includegraphics[width=0.4\textwidth]{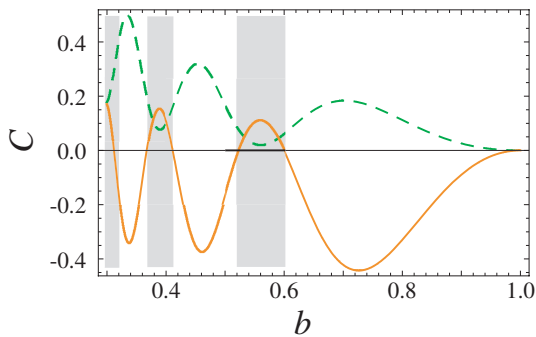}
\caption{(color online) The solid line denotes the expectation value of
$\{\hat{x}, \hat{p}\}$ in Eq.(\ref{eq:c0}) with the same parameters as its counterpart in Fig.\ref{fig-energy}. For comparison, the difference
in energy expectation between the two lines in Fig.\ref{fig-energy} is plotted in the dashed line. The gray areas highlight the regions when the wave-packet expands against the contraction of the trap. Although the energy input remains positive when this unexpected behavior occurs, its efficiency is greatly diminished due to the quantum friction. } \label{fig-c}
\end{figure}

\begin{figure}
\includegraphics[width=0.4\textwidth]{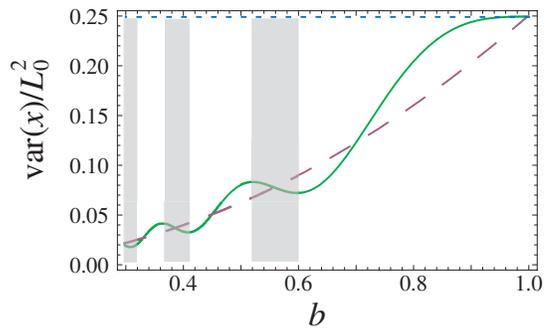}
\caption{(color online) The variance, var($x$)$=\langle x^{2}\rangle-\langle
x\rangle^{2}$, is plotted as a function of $b$ for $\dot {b}=-\omega_{0}/4$
(solid line), $-\omega_{0}/2000$ (dashed line and adiabatic limit), and
$-25\omega_{0}$ (dotted line and sudden limit). The length scale $L_{0}$ is defined as $(2\hbar/m\omega_{0})^{1/2}$. The gray area are the same regions of interest in Fig.\ref{fig-c}.} \label{fig-varx}
\end{figure}

Another useful signature of the wave-packet dynamics is the position-momentum correlation $\langle\hat{C}\rangle$ in
Eq.(\ref{eq:oscillationC}),
\begin{align}
\notag C(t)=&\hbar\frac{\dot{b}b}{\omega_{0}}\Big[\frac{1}{2}+\big(\frac{\dot{b}}{2\omega_{0}}\big)^{2}\Big]-\hbar\frac{b\dot{b}}{\omega_{0}}\Big[\big(\frac{\dot{b}}{2\omega_{0}}\big)^{2}+\frac{1}{2b}\Big]\cos\big(2\frac{\omega_{0}}{b}t\big)\\
&+\hbar b\big(\frac{\dot{b}}{2\omega_{0}}\big)^{2}\Big(\frac{1}{b}-2\Big)\sin\big(2\frac{\omega_{0}}{b}t\big),
\label{eq:c0}
\end{align} where $C(0)=0$ and a positive/negative $C(t)$ signifies an
expanding/contracting wave-packet. The numerical result in Fig.\ref{fig-c} reveals an unexpected motion; namely, the wave-packet can sometimes expand against the contracting trap. 
In Fig.\ref{fig-varx}, the variance var($x$)$=\langle
x^{2}\rangle-\langle x\rangle^{2}$ is calculated and plotted in the solid line. Consistent with the conclusion of Fig.\ref{fig-c}, the uncertainty in position also increases in the gray regions, in clear contrast to the monotonous adiabatic result in the dashed line.  

Since our choice of initial wavefunction $\phi_0(b=1)$ has nonzero projections on more than one direction of the orthonormal basis $\Psi_n(t,x)$ in Eq.(\ref{eq:wavefunc}), 
%is not a solution to the time-dependent Schr\"{o}dinger equation, 
the quantum breathing or the parasitic oscillation of its density profile is
expected. Similar to its classical counterpart, the quantum-friction term extracts work from  and thus \emph{slows down} the wave-packet when its motion is against
that of the potential profile. This resistance 
will modulate the energy expectation as well as its dynamics and result in oscillations coherently but not necessarily synchronized.
This is demonstrated by the contrast of the solid and dashed lines in Fig.\ref{fig-c}, which monitor the dynamics and the amount of energy input, respectively.

However, the harmonic oscillator is special because of its unique internal frequency and the classical analogies it enjoys, such as the dynamics of the coherent state.
In order to demonstrate (1) the authenticity of this oscillatory feature, (2) the dynamics and energy are coupled together with a common time lapse that is determinable, and (3) our approach of constructing a basis out of the original eigenfunctions is general to all potential traps, we shall examine the infinite potential well in the next section.

%However, cautions need to be taken at generalizing this seemingly intuitive classical example because it turns out that it has no quantum analogy\cite{{well}}. We believe this is due to the fact that, although the coupling between dynamics and the quantum friction still introduces oscillations in the energy, contributions from different modes cancel each other unless the potential exhibits a unique intrinsic frequency scale, such as the harmonic trap.

\section{Infinite well}
For any one-dimensional linearly-varying trap, the hamiltonian can be described by $\hat{H}(x/(b L_{0}))$ with the dimensionless factor $b(t)$ obeying $\dot{b}=$constant and $b(0)=1$. One convenient orthonormal basis can be chosen as
\begin{align}
\Phi_{n}(t,x)&=b^{-\frac{1}{2}}e^{-i\int_{0}^{t}dt'E_{n}/\hbar}
e^{im\dot{b}x^{2}/2b\hbar}\phi_{n}\Big(\frac{x}{bL_{0}}\Big).
\label{eq:general-wavefunc}
\end{align}
where $n$ are positive integers and $\phi_{n}(\frac{x}{b(t)L_{0}})$ and $E_{n}(t)$ are the instantaneous eigenstates and eigenenergies. 
%The parameter $b^{-1/2}$ is for time-varying normalization, 
The exponents, $(-\int_{0}^{t}dt'E_{n}(t')/\hbar)$ and $(m\dot{b}x^{2}/2b\hbar)$, provide the dynamic phases.
By using this basis, the energy expectation for any initial state $\psi_{0}$ can be found to share the same structure as Eq.(\ref{eq:packet-energy1}) with $\ddot{b}=0$.
Therefore, the arguments following Eq.(\ref{eq:packet-energy1}) can also be applied here, which allow us to conclude that the coupling between quantum friction and quantum breathing is general.

\begin{figure}
\includegraphics[width=0.4\textwidth]{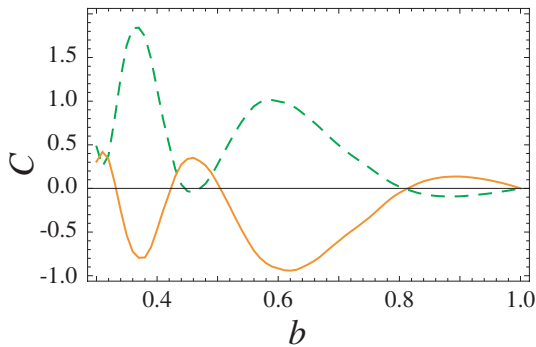}
\caption{(color online)The solid line denotes the expectation value of
$\{\hat{x}, \hat{p}\}$ with the $\dot{b}=E_{1}(0)/2\hbar$, $\beta=1/2E_{1}(0)$. As in Fig.\ref{fig-c}, the dashed line represents the energy expectation after subtracting the adiabatic contribution.}
\label{infinite-thermal-state}
\end{figure}

For illustration, let us pick an initial wave-packet in Boltzmann distribution of the instantaneous eigenstates at $t=0$:
\begin{align}
\psi_{0}(x)=N^{-1/2}\sum_{n=1} e^{-\beta E_{n}(0)}\phi_{n}(x/L_{0}),
\label{eq:period1}\end{align}
where $N=\sum$exp$(-\beta E_{n}(0))$.
Both the oscillation feature and the coupling between dynamic and energy are displayed clearly in the numerical result in Fig.\ref{infinite-thermal-state} for the operating temperature $\sim 1K$ accessible to cold-atom experiments.
Naively, one may not expect to see oscillations since there is no intrinsic frequency scale for an infinite well. 
It turns out that the phase $(n-l)\int_{0}^{t}dt'(\omega_{0}/b^{2})$ in Eq.(\ref{eq:packet-energy0}) can be reexpressed as $\int_{0}^{t}dt'(E_{n}-E_{l})/\hbar$. 
And this gives rise to a cutoff, $\int_{0}^{t}dt'(E_{1}-E_{2})/\hbar$, which defines a time lapse for each period
\begin{align}
T=\frac{h b(t)}{E_{2}(0)-E_{1}(0)}.
\label{eq:period}
\end{align}
This was checked to be consistent with the numerical results in Fig.\ref{infinite-thermal-state}. It comes as no surprise that we also find the above mentioned features persist in the limit of $\beta \rightarrow \infty$; namely, when the initial wavefunction is a pure state. In the mean time, we checked that the same oscillatory feature in the energy expectation was retained after the thermal averaging within the density matrix method.
%This reason why we chose Eq.(\label{eq:period1}) for our demonstration was that its curve was smoother.

\begin{figure}
\includegraphics[width=0.4\textwidth]{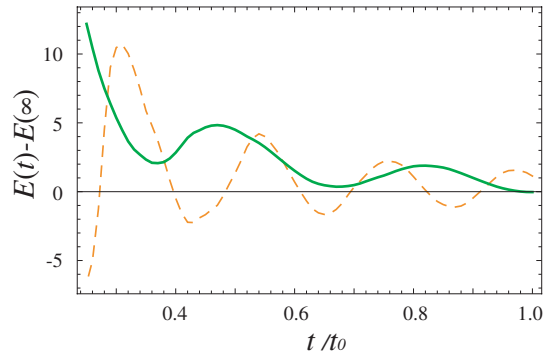}
\caption{(color online)The extra energy input in addition to the adiabatic result is shown for different contracting times in the unit $t_{0}=2\hbar/E_{1}(0)$. The final boundary is chosen arbitrarily to be at $b=0.3$ for the solid line and $b=0.2$ for the dashed line. Energies for the latter case have been suppressed by 10 times in order to fit in the plot. 
A short $t$ signifies a fast contraction. This oscillatory feature is reminiscent of that of RKKY coupling when plotted as a function of spatial distance.}
\label{exchange-coupling}
\end{figure}

For fixed initial and final boundary positions, we now examine the efficiency of different contracting rates for the energy input.
Figure \ref{exchange-coupling} shows the extra energy input compared to the adiabatic contribution oscillates with the contracting time. 
In contrast to the similar definition in Eq.(\ref{eq:period}), the time lapse $T$ is now a constant for a fixed $b$ and thus is a well-defined period.  
Since our derivations are applicable to all trapping systems by the choice of basis in Eq.(\ref{eq:general-wavefunc}),
we argue that the temporal oscillation in energy expectation with a fixed period is a common feature for all linearly-varying traps.
%To get a more comprehensive vision, we will compare the oscillatory feature in RKKY coupling and in inserting energy in next section. 

\section{Discussions and Conclusions}
%Oscillations in the energy input has an interesting connection with the indirect magnetic coupling in metals; namely, the interaction.
The moving boundaries are coupled through the atom trapped inside, and the energy input can be used to define the strength of this coupling. The oscillation with contracting time is reminiscent of the dependence on spatial distance between magnetic impurities in the Ruderman-Kittel-Kasuya-Yosida (RKKY) interaction mediated by the conducting electrons.
The coupling between magnetic side-layers in trilayer systems, such as the giant-magnetoresistence samples, has been shown\cite{chinghao} to be a simple extension of the RKKY formula: 
\begin{align}
\frac{\Delta E}{A}=2 {\rm Im}\int_{-\infty}^{E_{F}}dE\int \frac{d^{2}k_{\|}}{(2\pi)^2}\sum_{n=1}^{\infty}\frac{-1}{n}\big(r_{R}r_{L}e^{2i k_{\perp} D}\big)^{n},
\label{eq:ixc}
\end{align}
where $A$ is the interface area, $E_{F}$ is Fermi energy, $r_{R}/r_{L}$ represent the refraction coefficients in right/left interfaces, and  $D$ denotes the  spacer width. 
The perpendicular wavevector $k_{\perp}$ is a function of energy $E$ and parallel wavevector $k_{\|}$. 
The above equation indicates that the coupling energy is contributed by all possible scattering paths below the Fermi sea. Although Eq.(\ref{eq:ixc}) can not be solved analytically,  
numerical results show that it is an oscillatory function in $D$ with period $\pi/k_{F}$ which is half of the Fermi wavelength.
This period is expected because (1) the integrand consists of sinusoidal functions and (2) the Fermi energy defines a natural cutoff in the period\cite{md}.
In comparison, the oscillatory energy input in the time-varying traps is the result of quantum interference between different transition paths, and 
it also contains a well-defined period as $T=hb/E_{c}(0)$, as shown in Eq.(\ref{eq:period}). Here $E_{c}(0)$ plays the role of energy cutoff as the Fermi energy in the RKKY scenario.
Table \ref{tb} details the comparisons between these two oscillations. 
%It is clear that the inserting energy oscillation in particle-conservation system is a corresponding effect for the RKKY oscillation in many body system. 
\begin{table}[tbp]
\begin{tabular}[t]{|c|c| c|}
\hline
Oscillation & Energy Input & RKKY coupling\\
\hline\hline
Definition & $E(t)-E(t\rightarrow \infty)$ & $E(D)-E(D\rightarrow \infty)$ \\
Ensemble & Canonical & Grand Canonical\\
Contributors & Transition paths & Scattering paths\\
Parameter space & Contracting time & Spatial distance\\
Period &$ hb/E_{c}(0)$ & $\pi/k_{F}$\\
\hline
\end{tabular}
\caption{Comparison of quantum oscillations in the energy input for a time-varying trap and in the RKKY coupling.}
\label{tb}
\end{table}

In conclusion, we find that the energy expectation in time-varying traps is coupled to the dynamic fluctuations through quantum frictions.
Both the energy input and position-momentum correlation exhibit oscillatory features which we ascribe to the quantum interference among different transition paths. The time lapse in each period can be determined by the \emph{cutoff} in transition energies.
By fixing the initial and final trap profiles, we  find that 
the energy input is also an oscillatory function of the contracting time with an universal period $ hb/E_{c}(0)$ for all linearly-varying trap. Mathematically we show that the formalism of this problem is analogous to that of indirect magnetic coupling in metals. Therefore, the oscillation in the time domain of the time-varying trap shares the same physical origin as the RKKY oscillation in the spatial domain.

\appendix
\section{Harmonic Oscillator with a constant Adiabatic Parameter}
For a harmonic oscillator varying with a constant adiabatic parameter $\mu\equiv \dot{\omega}/\omega^{2}$, the time-dependent frequency is
\begin{align}
\omega(t)=\frac{\omega(0)}{1-\mu \cdot\omega(0) t}.
\end{align}
The phase factor can be solved as $\omega_{0}=\omega(0)\sqrt{1-\mu^{2}/4}$ by Eq.(\ref{eq:b-omega}).
It is obvious that the phase factor is only real for $|\mu|\le 2$.
The energy expectation is calculated via Eq.(\ref{eq:oscillationE}) as 
\begin{align}
\notag\langle \hat{H}(t) \rangle=&\notag\hbar\frac{\omega(t)}{\gamma}\Big\{ \Big[\frac{1}{2}+\langle\psi_{0}|\Lambda^{+}\Lambda|\psi_{0}\rangle\Big]\\
&+\frac{\mu}{2}\mathrm{Re}\Big[e^{-2i\gamma \theta(t)}(\mu/2+i\gamma)\langle\psi_{0}|\Lambda\Lambda|\psi_{0}\rangle\Big]\Big\},
\label{eq:mue}
\end{align}
where $\gamma=\sqrt{1-\mu^{2}/4}$ and $\theta(t)=\rm{ln}(\frac{\omega(t)}{\omega(0)})/\mu$.
The ratio between the oscillatory and non-oscillatory term is time-independent, which reveals that the energy expectation reduces to the adiabatic result 
when the time lapse passes each period of the oscillatory term.
For example, if we can select the ground state for the frequency $\omega(0)$ as the initial wave-packet, the energy expectation can be derived from Eq.(\ref{eq:mue}) 
%\begin{align}
%\notag \Lambda=&\sqrt{\gamma}\Big\{a^{+}(0)\big(1/2-1/2\gamma+i\mu/4\gamma \big)\\
%&+a(0)\big(1/2+1/2\gamma+i\mu/4\gamma \big)\Big\}
%\end{align}
\begin{align}
E(t)=\hbar\frac{\omega(t)}{2\gamma^{2}}\Big(1-\frac{\mu^{2}}{4}\cos\big(2\gamma \theta(t)\big)\Big).
\label{eq:groundmu}
\end{align}
The condition $2\gamma \theta(t_{n})=2n\pi$ defines the time lapses at which the energy returns to the adiabatic results $\hbar \omega(t)/2$ or, in other words, becomes frictionless.
These observations are consistent with those reported in Ref.\cite{mu}. Our Eq.(\ref{eq:groundmu}) is equivalent to their Eq.(8) after replacing the $2\gamma$ by $i\Omega$. 
In contrast to our approach that starts from the initial wavefuntion, 
Salamon and Rezek\cite{bang-bang,mu} estimated this energy by
connecting the expectaion values of Hamiltonian $\hat{H}$, Lagrangian $\hat{L}$ and correlation $\hat{C}$ for a time-dependent harmonic oscillator.
It will be heuristic to clarify the transition at $\mu =2$ by comparing our formulae, Eqs.(\ref{eq:oscillationE}) and (\ref{eq:oscillationC}), and Salamon and Rezek's in a future research.

\section{Adiabatic and Sudden limit}
 
To get a more comprehensive understanding of the oscillatory feature in time-varying traps, let us examine Eq.(\ref{eq:oscillationE}) in the adiabatic and sudden limits.
The competition between the time scales of the wave-packet and
the trap variation is best exemplified in the special case of a linear expansion or contraction,
$\dot{b}=v/L_{0}$ and $\ddot{b}=0$. The adiabatic
condition can be imposed by either (1)
$|\sqrt{2}\dot{\omega}/(8\omega^{2})|\ll 1$, which states that the energy input
from the time-varying trap is much smaller than the
intrinsic energy of the wave-packet\cite{optimal} or the more direct (2) $|v/v_{p}|\ll 1$ which restricts the potential to varying much slower than the averaged atomic speed, $v_{p}=
\sqrt{2}\hbar/(mbL_{0})$, in a trap with an intrinsic angular frequency, $\omega(t)=\omega_{0}/b^{2}$. Both
approaches guarantee that $(b\dot{b}/\omega_{0})\ll
1$ and Eq.(\ref{eq:oscillationE}) can be simplified to give the energy in the adiabatic approximation as
\begin{align}
\langle \hat{H}(t) \rangle\approx
\frac{\hbar\omega_{0}}{b^{2}}\big(N+\frac{1}{2}\big)=\frac{\langle
\hat{H}(0) \rangle}{b^{2}}, \label{eq:adabaticE}
\end{align}
where $N=\langle a^{+}a\rangle$.
The oscillation term can be neglected in this limit.

The sudden approximation describes the other limit of $(b\dot{b}/\omega_{0})\gg 1$ when the
atomic motion can not catch up with the trap variation. The Taylor expansion works
for the exponential term in Eq.(\ref{eq:oscillationE}) at this limit
\begin{align}
e^{-2i\int_{0}^{t}dt'\omega_{0}/b^{2}}=\sum_{n=0}\frac{1}{n!}
\Big[2i\frac{\omega_{0}}{b\dot{b}}\Big(1-b\Big)\Big]^{n}.
\label{eq:suddentaylor}
\end{align}
Keep the expansion up to the fourth order and insert to
Eq.(\ref{eq:oscillationE}) while retaining the lowest order term in
$\omega_{0}/(b\dot{b})$ give the energy expectation as
\begin{align}
\notag\langle \hat{H} \rangle&\approx
\hbar\omega_{0}\big(N+\frac{1}{2}\big)+\frac{\hbar\omega_{0}}{4}\big(\frac{1}{b^4}-1\big)\langle(a+a^{+})^{2}\rangle\\
&=\langle \hat{H}(0) \rangle+
\frac{1}{2}\big(\omega^{2}(t)-\omega^{2}_{0}\big)\langle
x^{2}\rangle \label{eq:suddenperturb}
\end{align}
with $\omega(t)=\omega_{0}/b^{2}$. This is the same result as is obtained by the time-independent perturbation theory. It is worth mentioning that, although this conclusion is independent of the sign of $\dot{b}$ for a harmonic
potential, the sudden approximation will break down terribly when applied to a contracting infinite well because of the shrinkage of Hilbert space. 
Strictly speaking, the oscillatory term that we highlighted in Eq.(\ref{eq:oscillationE}) does not show up in the adiabatic and sudden limits for different reasons. In the former, it is suppressed due to the smallness of the coefficient in Eq.(\ref{eq:oscillationE}), while the Taylor expansion of Eq.(\ref{eq:suddentaylor}) guarantees its minuteness in the latter.
Finally, note that the energy expectation does not depend on $\dot{b}$ explicitly in these two limits.

\end{document}